\documentclass[a4paper]{cas-dc}
\usepackage[numbers,sort&compress]{natbib}
\usepackage{braket}
\usepackage{multicol}
\usepackage{subfigure}

\definecolor{darkpastelgreen}{rgb}{0.01, 0.75, 0.24}
\definecolor{electricindigo}{rgb}{0.44, 0.0, 1.0}
\definecolor{palatinateblue}{rgb}{0.15, 0.23, 0.89}
\definecolor{carminered}{rgb}{1.0, 0.0, 0.22}
\hypersetup{linkcolor=palatinateblue, citecolor=electricindigo, urlcolor=carminered}

\begin{document}
\shorttitle{Lifshitz formulas for finite-density Casimir effect}
\shortauthors{D. Fujii, K. Nakayama and K. Suzuki}  
\title[mode = title]{Lifshitz formulas for finite-density Casimir effect}  

\author[1,2]{Daisuke Fujii}[type=editor, orcid=0000-0002-6298-9278]
\nonumnote{$\dag$ E-mail: daisuke@rcnp.osaka-u.ac.jp (corresponding author)}

\author[3]{Katsumasa Nakayama}[orcid=0000-0003-0270-8523
]
\nonumnote{$\ddag$ E-mail: katsumasa.nakayama@riken.jp
(corresponding author)}

\author[1]{Kei Suzuki}[orcid=0000-0002-8746-4064]
\nonumnote{$\ddag$ E-mail: k.suzuki.2010@th.phys.titech.ac.jp (corresponding author)}

\address[1]{Advanced Science Research Center, Japan Atomic Energy Agency (JAEA), Tokai, 319-1195, Japan}
\address[2]{Research Center for Nuclear Physics, Osaka University, Ibaraki 567-0048, Japan}
\address[3]{RIKEN Center for Computational Science, Kobe, 650-0047, Japan}

\begin{abstract}
The Lifshitz formula is well known as a theoretical approach to investigate the Casimir effect at finite temperature.
In this Letter, we generalize the Lifshitz formula to the Casimir effect originating from quantum fields at finite chemical potential.
To demonstrate the versatility of this formula, we discuss the typical phenomena of the Casimir effect at finite chemical potential in various systems, such as some boundary conditions, finite temperatures, arbitrary spatial dimensions, and mismatched chemical potentials.
This formula can be applied to the Casimir effect in dense quark matter and Dirac/Weyl semimetals, where the chemical potential is regarded as a parameter to control the Casimir effect.
\end{abstract}


\maketitle

\section{Introduction}
In 1948, Casimir predicted that the zero-point energy of photon fields in vacuum sandwiched by two perfectly conducting parallel plates induces an attractive force~\cite{Casimir:1948dh}, which is the so-called {\it Casimir effect} (see Refs.~\cite{Lamoreaux:1996wh,Bressi:2002fr} for experiments and Refs.~\cite{Plunien:1986ca,Mostepanenko:1988bs,Bordag:2001qi,Milton:2001yy,Klimchitskaya:2009cw,Woods:2015pla,Gong:2020ttb,Lu:2021jvu} for reviews).
After that, in 1956, Lifshitz derived an alternative formula~\cite{Lifshitz:1956zz} which is nowadays called the {\it Lifshitz formula}.
This is a regularization technique to remove the divergence of the zero-point energy and reproduce Casimir's original result.
Furthermore, this formula predicts the dependence on finite temperature and/or dielectric functions of parallel plates.
The study of the Casimir effect at finite temperature is practically needed because 
experiments for the measurement of Casimir force are, more or less, exposed to a finite-temperature environment.

Whereas the {\it thermal Casimir effect}\footnote{Precisely speaking, ``thermal'' for the Casimir effect has two meanings: (i) temperature dependence of the free energy for photon fields and (ii) temperature dependence of the dielectric functions of materials composing of boundary conditions such as parallel plates.} was well-established by the excellent agreement between theory~\cite{Lifshitz:1956zz,Fierz:1960zq,Mehra:1967wf,Brown:1969na,Schwinger:1977pa} and experiment~\cite{Sushkov:2010cv}, the counterpart at finite chemical potential is still unknown.
The main reason is that the conventional study of the Casimir effect focuses on the photon field, and it is usually difficult to control the chemical potential of photons in equilibrium (for non-equilibrium cases, see Refs.~\cite{Chen:2016,Spreng:2024oet}).
On the other hand, if one focuses on the Casimir effect originating from fermion fields~\cite{Johnson:1975zp,Mamaev:1980jn,Gundersen:1987wz}, the corresponding chemical potential can be a significant parameter to control the Casimir effect:
the fermionic counterpart of the Casimir effect might be realized in quark systems and Dirac/Weyl materials, but its experimental observation is still an open problem, which requires more controllable parameters.
Theoretically, an approach that correctly implements chemical potentials is needed, but it has not yet been established.
In this Letter, for the first time, we extend the Lifshitz formula to systems at finite chemical potential.

\section{Main formula}

First, we show our main finding.
As a typical example, we consider the Dirac field with a mass $M$ and at chemical potential $\mu$ in the (3+1) dimensional spacetime.
As an idealized boundary condition, we impose the periodic boundary conditions (PBCs) at $z=0$ and $z=L_z$ in the $z$ direction, where the momentum is discretized as $k_z\to 2n\pi/L_z$ $(n=0,\pm1,\pm2,\cdots)$.
Then, the Lifshitz formula for the Casimir energy $E_\mathrm{Cas}$ (per unit area) is written as
\begin{align}
E_\mathrm{Cas} &= -4 \int_{-\infty}^\infty \frac{d\xi}{2\pi} \int \frac{dk_xdk_y}{(2\pi)^2} \ln \left[1-e^{-L_z \tilde{k}_z} \right], \label{eq:ECas_PBC}\\
\tilde{k}_z &= \sqrt{M^2+ k_x^2 +k_y^2  -(i\xi + \mu)^2},
\end{align}
in the natural unit of $\hbar=c=1$ with the reduced Planck constant $\hbar$ and the speed of light $c$.
Here, $\xi$ is the imaginary part of the imaginary frequency $i\xi$, and $k_x$ and $k_y$ are the momenta in the perpendicular direction.
The overall factor of $-2$ is regarded as the minus sign from the fermion zero-point energy and the spin degrees of freedom.
The remaining factor of $2$ and the form of $1-e^{-L_z \tilde{k}_z}$ depend on the boundary condition and are now characterized by the PBC.
By substituting $\mu=0$ to this formula, we obtain the conventional Lifshitz formula.

\begin{figure}[t!]
    \centering
    \begin{minipage}[t]{1.0\columnwidth}
    \includegraphics[clip,width=1.0\columnwidth]{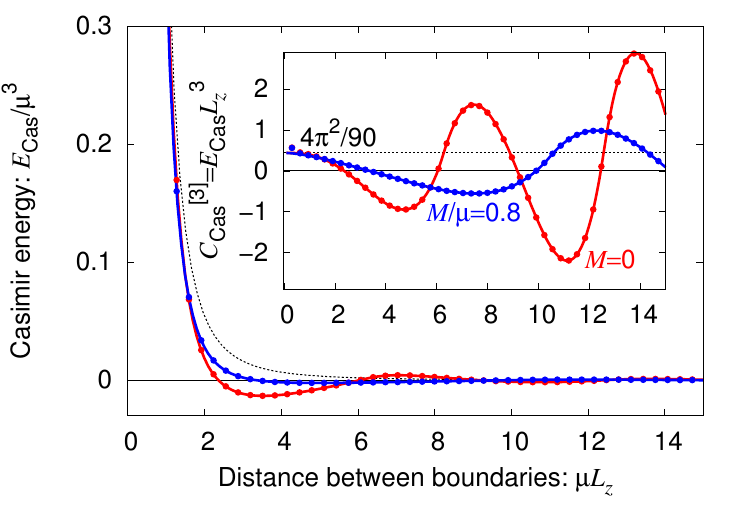}
    \end{minipage}
    \caption{Typical behaviors of the Casimir energy $E_\mathrm{Cas}$ and its coefficient $C_\mathrm{Cas}^{[3]}$ at finite chemical potential larger than the mass ($\mu >M$).
Solid lines: from the Lifshitz formula.
Points: from the lattice regularization.
The dotted line represents $E_\mathrm{Cas} = 4\pi^2/90L_z^3$ known for the massless field at $\mu=0$.}
    \label{fig:3d}
\end{figure}

Using Eq.~(\ref{eq:ECas_PBC}), in Fig.~\ref{fig:3d}, we show a typical behavior of the Casimir energy at finite chemical potential, 
Here, we defined a Casimir coefficient $C_\mathrm{Cas}^{[3]} \equiv E_\mathrm{Cas} \times L_z^3$ in order to visualize the $1/L_z^3$ scaling well known for massless fields.
For $\mu \leq M$ (i.e., at zero density), we obtain the conventional behavior of the Casimir effect for massless or massive fields because there is no contribution from the Fermi sea. 
For $\mu>M$ (i.e., at finite density), we find an oscillation of the Casimir energy as a function of $L_z$.
This oscillation is caused by the relationship between the fixed Fermi level and the $L_z$-dependent eigenvalues discretized by boundary conditions (for a graphical explanation, see Ref.~\cite{Fujii:2024fzy}).
For this reason, the oscillation period is given as,
\begin{align}
L_z^\mathrm{osc}= \frac{2\pi}{\sqrt{\mu^2-M^2}}. \label{eq:Lzosc}
\end{align}
Using this formula, we get $L_z^\mathrm{osc}=2\pi/\mu$ for the massless field, and the massive field leads to a longer period.
Note that the so-called {\it oscillating Casimir effect} (or  Casimir-like interaction) can be induced by various types of systems and origins (e.g., see Refs.~\cite{Bulgac:2001np,Fuchs:2007,Wachter:2007,Kolomeisky:2008,Nishida:2008ra,Zhabinskaya:2009,Chen:2011xtv,Chen:2011JStatPhys,Kruger:2011zd,Kruger:2012bh,Han:2015,Chen:2016,Jiang:2018ivv,Ishikawa:2020ezm,Ishikawa:2020icy,Mandlecha:2022cll,Nakayama:2022fvh,Nakata:2023keh,Fujii:2024fzy,Spreng:2024oet}), but Eq.~(\ref{eq:ECas_PBC}) is regarded as a new formula containing both the non-oscillating Casimir effect at $\mu=0$ and the oscillating one at $\mu \neq 0$. 

To check the validity of the Lifshitz formula~(\ref{eq:ECas_PBC}), we compare with the results obtained by the lattice regularization approach~\cite{Actor:1999nb,Pawellek:2013sda,Ishikawa:2020ezm,Ishikawa:2020icy,Nakayama:2022ild,Nakata:2022pen,Mandlecha:2022cll,Nakayama:2022fvh,Swingle:2022vie,Nakata:2023keh,Flores:2023whr,Nakayama:2023zvm,Beenakker:2024yhq} (at finite $\mu$, see Appendix~\ref{App:lattice} or Ref.~\cite{Fujii:2024fzy}).
The lattice regularization can usually reproduce the correct result by taking the continuum limit ($a\to0$) whereas at a nonzero lattice spacing ($a>0$) the result in the short-$L_z$ region deviates.
In Fig.~\ref{fig:3d}, we fix the lattice spacing as $\mu a =0.08$ which is small enough.
We can see that, in the longer-$L_z$ region, both the results well coincide, which suggests that Eq.~(\ref{eq:ECas_PBC}) describes the correct behaviors of the Casimir effect.

\section{Derivation}

Here, we overview a derivation of the Lifshitz formula~(\ref{eq:ECas_PBC}).
We start from the zero-point energy (i.e., the sum of eigenvalues) of the Dirac field under the PBC and finite chemical potential:
\begin{align}
    &-2\sum^\infty_{n=-\infty}\frac{(|\omega_{n}|+|\tilde{\omega}_n|)}{2}=-2 \sideset{}{^\prime}{\sum}^\infty_{n=0} (|\omega_{n}|+|\tilde{\omega}_n|), \\
    &\omega_n=E-\mu, \ \ \ \tilde{\omega}_n=-E-\mu, \notag \\
    &E=\sqrt{M^2 + k_x^2+k_y^2+(2n \pi/L_z)^2}. 
\end{align}
The overall factor of $-2$ is from the fermion statistics and the spin degrees of freedom.
The prime in the sum means that the factor $1/2$ is multiplied only for $n=0$.
Note that we omit the integrals with respect to $k_x$ and $k_y$ for a moment, but it will be restored in the final form.
By the argument principle\footnote{The derivation using the argument principle is standard also at zero chemical potential~\cite{Kampen:1968,Schram:1973,Bordag:2001qi}.}, using an integer $n_{\rm F}=\lfloor L_z k_{\rm F}/2\pi\rfloor$ with the Fermi momentum $k_{\rm F}$, the infinite sum can be evaluated as
\begin{align}
    & \sideset{}{^\prime}{\sum}^{n_{\rm F}}_{n=0} (-\omega_{n})+\sideset{}{^\prime}{\sum}^\infty_{n=0} (-\tilde{\omega}_n)=\frac{1}{2\pi i}\oint_{C} \omega d\ln\Delta_-(\omega), \\ &\sideset{}{^\prime}{\sum}^\infty_{n=n_{\rm F}+1} \omega_{n}=\frac{1}{2\pi i}\oint_{C} \omega d\ln\Delta_+(\omega), \\ 
    &\frac{1}{2\pi i}\oint_C \omega d\ln\Delta_\pm(\omega)=\frac{1}{2\pi i}\left(\int^{-i\infty}_{i\infty}+\int_{C_+}\right)\omega d\ln\Delta_\pm(\omega). \label{Contour}
\end{align}
The contour integral along the closed path $C$ on the complex $\omega$ plane consists of the infinite integral on the imaginary axis and the counterclockwise integral along $C_+$ on a semicircle (in the right half of the $\omega$ plane) with an infinite radius centered at the origin.
$\Delta_\pm(\omega)$ is a meromorphic function with no pole
\begin{align}
    &\Delta_\pm(\omega)=1-e^{-ik^{[\pm]}_zL_z}, \label{delta} \\    &ik^{[\pm]}_z \equiv \tilde{k}^{[\pm]}_z =\sqrt{M^2 + k^2_x+k^2_y-(\omega\pm\mu)^2}. \notag
\end{align}
The zero points of $\Delta_+(\omega)$ in the region surrounded by $C$ are $\omega=\omega_n$ at $n \geq n_{\rm F}+1$ (i.e., the eigenvalues higher than the Fermi level) on the real axis.
On the other hand, the zero points of $\Delta_-(\omega)$ in $C$ contain $\omega=\tilde{\omega}_n$ as well as $\omega=\omega_n$ at $n\leq n_{\rm F}$ (i.e., the eigenvalues lower than the Fermi level). 

With the imaginary frequency $\omega\equiv i\xi$, the first term of Eq.~\eqref{Contour} vanishes because of $\xi d\ln\Delta_s(i\xi)\rightarrow0$ in the limit of $\xi\rightarrow\infty$. 
Furthermore, by performing integration by parts on the second term of Eq.~\eqref{Contour}, we obtain the Lifshitz formula. 
Finally, because $\pm\mu$ in Eq.~(\ref{delta}) leads to the same result, we obtain the formula \eqref{eq:ECas_PBC}.

\section{Applications}

\begin{figure}[t!]
    \centering
    \begin{minipage}[t]{1.0\columnwidth}
    \includegraphics[clip,width=1.0\columnwidth]{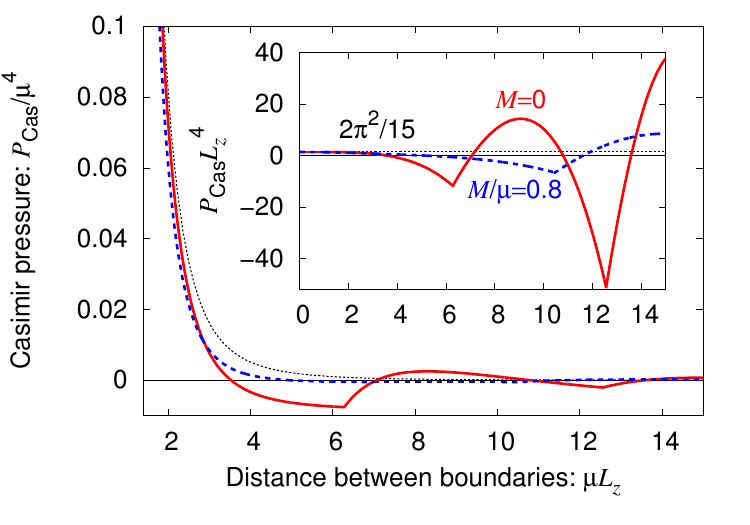}
    \end{minipage}
    \caption{Typical behaviors of the Casimir pressure $P_\mathrm{Cas}$ and its coefficient $P_\mathrm{Cas}L_z^4$ at finite chemical potential larger than the mass ($\mu >M$).
The dotted line represents $P_\mathrm{Cas} = 2\pi^2/15L_z^4$ known for the massless field at $\mu=0$.}
    \label{fig:pressure}
\end{figure}

{\it Application 1: Casimir pressure and Casimir force.}--- Experimentally, when the measurement of the energy difference is difficult, more realistic observables are the Casimir pressure and Casimir force.
These quantities can be easily obtained from the $L_z$ derivative of the Casimir energy~(\ref{eq:ECas_PBC}):
\begin{align}
P_\mathrm{Cas} &\equiv -\frac{dE_\mathrm{Cas}}{dL_z} = 4 \int_{-\infty}^\infty \frac{d\xi}{2\pi} \int \frac{dk_xdk_y}{(2\pi)^2} \frac{\tilde{k}_z}{e^{L_z \tilde{k}_z}-1},\\
F_\mathrm{Cas} &\equiv L_xL_y P_\mathrm{Cas} = - L_xL_y \frac{dE_\mathrm{Cas}}{dL_z}.
\end{align}
These can be called the Lifshitz formula for Casimir pressure and force at finite $\mu$.
In Fig.~\ref{fig:pressure}, we show a typical behavior of the Casimir pressure.
We find that, similar to the Casimir energy, the corresponding pressure and force also oscillate, but their waveforms are different from that of the Casimir energy.
In experiments, when a material is extremely thin, the Casimir pressure could stretch or compress the material (depending on the compressibility or elastic modulus of the material), and may be significant as a measurable quantity.

{\it Application 2: Boundary conditions.}---The cases with the other boundary conditions can be derived in a similar manner.
For the antiperiodic boundary conditions (APBC), the discrete momentum is given as $k_z\to (2n+1)\pi/L_z$.
Then, the Lifsthiz formula is
\begin{align}
E_\mathrm{Cas}^\mathrm{APBC} &= -4 \int_{-\infty}^\infty \frac{d\xi}{2\pi} \int \frac{dk_xdk_y}{(2\pi)^2} \ln \left[1+e^{-L_z \tilde{k}_z} \right]. \label{eq:ECas_APBC}
\end{align}
The difference from Eq.~(\ref{eq:ECas_PBC}) is only the sign in front of $e^{-L_z \tilde{k}_z}$.
Note that, the oscillation period is the same as Eq.~(\ref{eq:Lzosc}) for the PBC.

Similarly, we can obtain the formula for the boundary conditions leading to $k_z\to (n+1/2)\pi/L_z$ $(n=0,1,2,\cdots)$, which corresponds to the MIT bag boundary conditions~\cite{Chodos:1974je} well known for massless Dirac fields:
\begin{align}
E_\mathrm{Cas}^\mathrm{MIT} &= -2 \int_{-\infty}^\infty \frac{d\xi}{2\pi} \int \frac{dk_xdk_y}{(2\pi)^2} \ln \left[1+e^{-2L_z \tilde{k}_z} \right]. \label{eq:ECas_MIT}
\end{align}
The differences from Eq.~(\ref{eq:ECas_APBC}) are the overall factor $1/2$ and the exponential function $e^{-2L_z \tilde{k}_z}$.
For the latter reason, the oscillation period is shorter than Eq.~(\ref{eq:Lzosc}) for the PBC/APBC by a factor of $1/2$: $L_z^\mathrm{osc} = \pi/\sqrt{\mu^2-M^2}$.
Thus, our formulas can be applied to various boundary conditions.\footnote{Another famous boundary condition is the case leading to the discrete momentum $k_z\to n\pi/L_z$.
This situation is realized by the Dirichlet or Neumann boundaries for scalar fields and by perfectly conducting plates for photon fields.
The corresponding formula is obtained by putting the minus sign in front of $e^{-2L_z \tilde{k}_z}$ in Eq.~(\ref{eq:ECas_MIT}) (and by multiplying by appropriate factors for specific quantum fields).}

\begin{figure}[t!]
    \centering
    \begin{minipage}[t]{1.0\columnwidth}
    \includegraphics[clip,width=1.0\columnwidth]{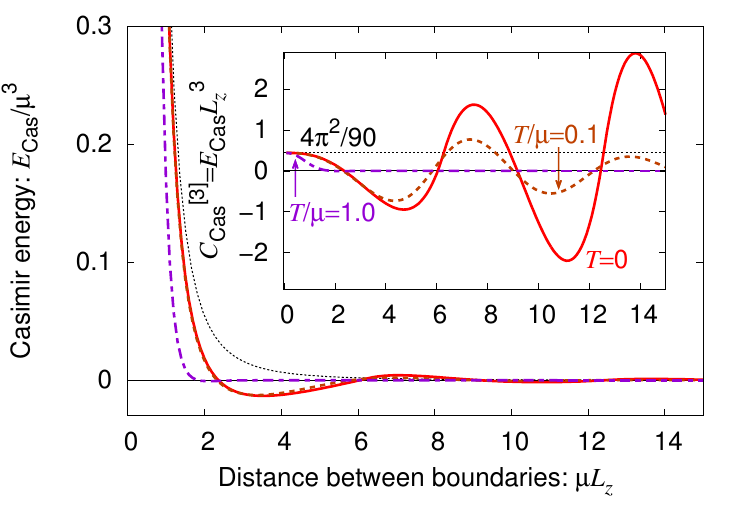}
    \end{minipage}
    \caption{
Temperature dependence ($T/\mu=0,0.1,1.0$) of Casimir energy $E_\mathrm{Cas}$ and its coefficient $C_\mathrm{Cas}^{[3]}$ from massless Dirac fields at finite chemical potentials $\mu$.}
    \label{fig:temp}
\end{figure}

{\it Application 3: Finite temperature.}---At finite temperature, we replace the infinite integral $\int_{-\infty}^\infty \frac{d\xi}{2\pi} f(\xi)$ with the infinite sum $\sum_{l=-\infty}^\infty f(\xi_l)$, where $\xi_l =(2 l+1) \pi T$ is the fermion Matubara frequency with the label $l=0,\pm1,\cdots$ and the temperature $T$.
Using this replacement, Eq.~(\ref{eq:ECas_PBC}) is deformed as
\begin{align}
E_\mathrm{Cas} (T) &= -4T \sum_{l=-\infty}^\infty \int \frac{dk_xdk_y}{(2\pi)^2} \ln \left[1-e^{-L_z \tilde{k}_z} \right], \\
\tilde{k}_z &= \sqrt{ M^2 +k_x^2 +k_y^2 -(i\xi_l + \mu)^2}. \nonumber
\end{align}
This formula contains all the $L_z$ dependences from the vacuum, finite $\mu$, and finite $T$.
The $T\to 0$ limit is equivalent to Eq.~(\ref{eq:ECas_PBC}).\footnote{Also, the $\mu\to 0$ limit is consistent with the known formula (see, e.g., Ref.~\cite{Gundersen:1987wz} for the fermionic Casimir effect at finite temperature).}
In Fig.~\ref{fig:temp}, we compare the results at $T=0$, $0.1\mu$, and $1.0\mu$.
We find a suppression of the Casimir effect due to the temperature.\footnote{
Note that the temperature dependence of the Casimir effect depends on boson or fermion fields.
Since we now consider the Dirac fermion field, the total (i.e., zero-temperature plus finite-temperature) Casimir effect is suppressed as a function of temperature.
On the other hand, the bosonic thermal Casimir effect is usually enhanced.}

\begin{figure}[tb!]
    \centering
    \begin{minipage}[t]{1.0\columnwidth}
    \includegraphics[clip,width=1.0\columnwidth]{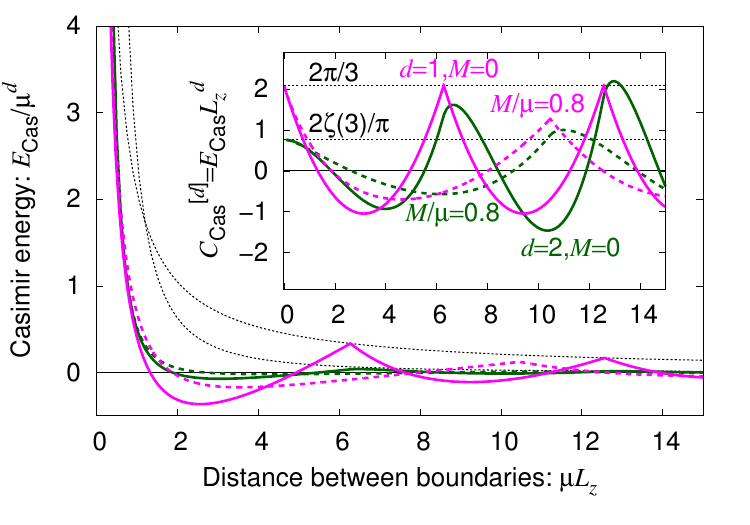}
    \end{minipage}
    \caption{
Casimir energy $E_\mathrm{Cas}$ and its coefficient $C_\mathrm{Cas}^{[d]}$ at spatial dimension $d=2$ and $d=1$.
The dotted lines represent $E_\mathrm{Cas} = 2\zeta(3)/\pi L_z^2$ and $2\pi/3L_z$ known for the massless field at $\mu=0$.}
    \label{fig:dim}
\end{figure}

{\it Application 4: Spatial dimensions.}---The formula~(\ref{eq:ECas_PBC}) is in the (3+1) dimensional spacetime, but it can be generalized to arbitrary ($d$+1) dimensional spacetime by replacing as the transverse momenta $dk_x dk_y/(2\pi)^2 \to dk_{x_1} dk_{x_2} \cdots dk_{x_d}/(2\pi)^d$ and $k_x^2+k_y^2 \to k_{x_1}^2 +k_{x_2}^2 + \cdots + k_{x_d}^2$. 
Then,
\begin{align}
E_\mathrm{Cas}(d)&= -4 \int_{-\infty}^\infty \frac{d\xi}{2\pi} \int \frac{dk_{x_1}dk_{x_2}\cdots}{(2\pi)^d} \ln \left[1-e^{-L_z \tilde{k}_z} \right], \label{eq:d_dep}\\
\tilde{k}_z &= \sqrt{M^2 + k_{x_1}^2 +k_{x_2}^2 +\cdots + k_{x_d}^2  - (i\xi + \mu)^2},
\end{align}
where we kept the spin degrees of freedom $2$ for simplicity.
In particular, the lower dimensional ($d=2$ or $1$) systems are realized in condensed matter physics.
In Fig.~\ref{fig:dim}, we show the typical behaviors of the Casimir energies and coefficients $C_\mathrm{Cas}^{[d]} \equiv E_\mathrm{Cas} \times L_z^d$.
For both $d=1$ and $2$, we find oscillatory behaviors, and their periods are characterized by $2\pi/\sqrt{\mu^2-M^2}$: the period is independent of $d$.
At $d=1$, $E_\mathrm{Cas}$ is nondifferentiable at a certain $L_z$, which means that $P_\mathrm{Cas}$ is discontinuous at the same $L_z$.
This is different from $d=2,3$, where $P_\mathrm{Cas}$ is nondifferentiable (see Fig.~\ref{fig:pressure} at $d=3$).
Furthermore, we find that the $\mu$-dependent part of $E_\mathrm{Cas}$ is scaled as $1/L_z^{(d+1)/2}$, which is distinct from the vacuum part scaled as $1/L_z^d$ at $M=0$.
Thus, the dimensional structure of the Fermi sea (i.e., Fermi sphere, Fermi circle, and Fermi line) characterizes the typical behavior of the oscillating Casimir effect.
Conversely, the measurement of the oscillatory behavior is useful as a signal of the dimensional structures of the dominant quantum fields.

\begin{figure}[t!]
    \centering
    \begin{minipage}[t]{1.0\columnwidth}
    \includegraphics[clip,width=1.0\columnwidth]{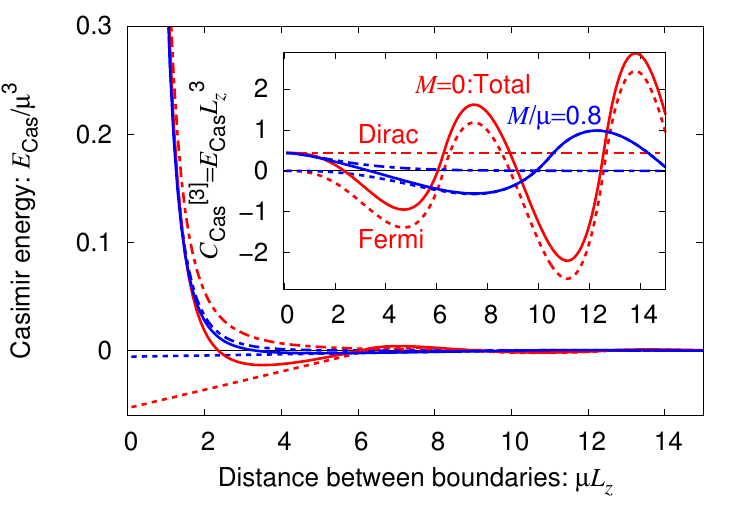}
    \end{minipage}
    \caption{
Contributions of Dirac and Fermi seas for Casimir energy $E_\mathrm{Cas}$ and its coefficient $C_\mathrm{Cas}^{[3]}$.}
    \label{fig:separation}
\end{figure}

{\it Application 5: Separation of Dirac and Fermi seas.}--- While Eq.~(\ref{eq:ECas_PBC}) contains both the contributions from the zero-point energy from the $\mu$-independent vacuum (the Dirac sea) and the $\mu$-dependent energy (the Fermi sea), we can obtain only that from the Dirac sea by substituting $\mu=0$.
Therefore, we can also calculate only the contribution from the Fermi sea by subtracting the Dirac-sea contribution $E_\mathrm{Cas}(\mu=0)$ from the total Casimir energy $E_\mathrm{Cas}(\mu)$: 
\begin{align}
E_\mathrm{Cas}^\mathrm{Fermi} = E_\mathrm{Cas}(\mu) -  E_\mathrm{Cas}(\mu=0).
\end{align}
Thus, using the combination of the Lifshitz formulas, we can describe even the contribution of the Fermi sea.
In Fig.~\ref{fig:separation}, we compare the total, Dirac-sea, and Fermi-sea contributions.
Thus, the Casimir effect in Eq.~(\ref{eq:ECas_PBC}) correctly includes the contribution from the Dirac sea, which is regarded as a ``bare" fermionic Casimir effect.

\begin{figure}[b!]
    \centering
    \begin{minipage}[t]{1.0\columnwidth}
    \includegraphics[clip,width=1.0\columnwidth]{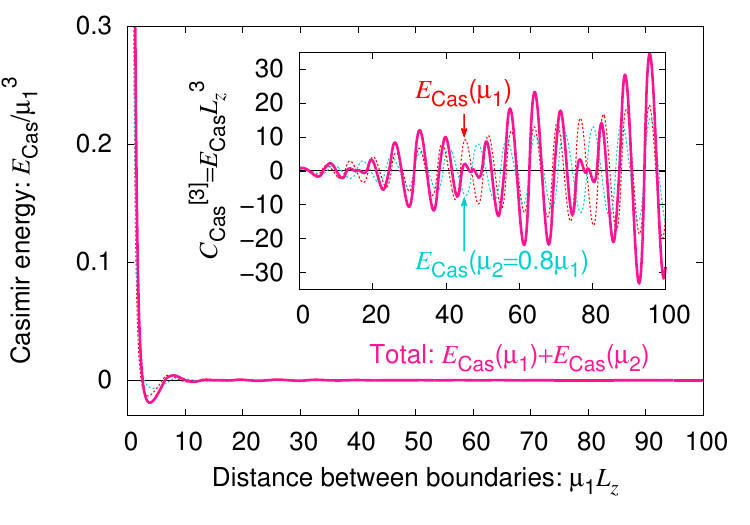}
    \end{minipage}
    \caption{
Casimir energy $E_\mathrm{Cas}$ and its coefficient $C_\mathrm{Cas}^{[3]}$ from two massless Dirac fermion fields at mismatched chemical potentials $\mu_1$ and $\mu_2$.}
    \label{fig:beat}
\end{figure}

{\it Application 6: Mismatched chemical potentials.}---For two kinds of Dirac fields at different chemical potentials $\mu_1$ and $\mu_2$, the total Casimir energy is represented as the sum of the two Casimir energies:
\begin{align}
E_\mathrm{Cas} (\mu_1,\mu_2) = E_\mathrm{Cas}(\mu_1) + E_\mathrm{Cas}(\mu_2).
\end{align}
If $\mu_1= \mu_2$, then the total Casimir energy is twice as large as one Casimir energy.
If $\mu_1\neq \mu_2$, the total Casimir energy oscillates as a superposition of the two oscillations, and a new period appears.
This is the so-called {\it beating Casimir effect}.\footnote{This beating Casimir effect is a new phenomenon in the sense that it arises from the multiple chemical potentials.
Similar beating Casimir effects were predicted from other origins: spin-split Dirac points~\cite{Nakayama:2022fvh} and multiple exceptional points~\cite{Nakata:2023keh}.}
As an example, in Fig.~\ref{fig:beat}, we show the results for the two massless Dirac fields at $\mu_1$ and $\mu_2=0.8\mu_1$.
Then, the period of the beat is estimated as $1/L_z^\mathrm{beat}=\mu_1/2\pi-\mu_2/2\pi$: $\mu_1 L_z^\mathrm{beat} \sim 31.4$.
Note that mismatching of two (or more) chemical potentials is not a rare situation and is frequently realized in nuclear physics: the chemical potentials of protons and neutrons in nuclear matter (or up and down quarks in quark matter) are different in environments such as neutron stars and neutron-rich nuclei.

\section{Physical examples}

Finally, we show some examples of physical platforms for our formulas. 

{\it Physical example 1: Quark matter.}---We emphasize that Eq.~(\ref{eq:ECas_PBC}) is a standard formula for the Dirac field at finite chemical potential, and such a situation is realistic for quark fields in dense quark matter.
Since the free up and down quarks are approximately massless and their velocity is close to the speed of light, the magnitude of the Casimir energy can be comparable with the photonic one and is enhanced by the factors of flavors and colors.
In particular, a thin domain of dense quark matter~\cite{Fujii:2024fzy} or quark-gluon plasma~\cite{Hashimoto:1984zz,Jackson:1986nc,Saito:1990jr} is regarded as the Casimir effect-like geometry.
Dynamics of interacting quarks is described by quantum chromodynamics (QCD).
Its nonperturbative analysis is difficult, but effective models based on a quasiparticle picture of quarks can be utilized. 
Our formulas will be useful in various types of effective models of quasiquarks.
The Casimir effect in dense QCD (or effective models of dense quark matter) will be preciously examined by numerical lattice simulations (free from the sign problem).
Recently, because lattice simulations of the Casimir effect for Yang-Mills fields are vigorously developed~\cite{Chernodub:2018pmt,Chernodub:2018aix,Kitazawa:2019otp,Chernodub:2023dok}, the comprehensive study including quark fields is an urgent issue. 

{\it Physical example 2: Dirac/Weyl semimetals.}---Similar to quark matter, (three-dimensional) Dirac or Weyl semimetals~\cite{Armitage:2017cjs} are another testing ground for the Casimir effect originating from Dirac or Weyl electron fields.
In particular, thin films of these materials are regarded as the Casimir effect-like geometry~\cite{Nakayama:2022fvh}.
Then, the typical magnitude of the fermionic Casimir effect is characterized by the Fermi velocity (typically, $0.1$-$1$\% of the speed of light), but its contribution is relevant as a part of thermodynamics inside materials.
Experimentally, the chemical potential (i.e., the position of the Fermi level) in these materials can be controlled by doping electrons~\cite{Liu:2014} or applying gate voltages~\cite{Liu:2015}, which means that the $\mu$-dependent Casimir effect can also be controlled.
Our formulas will be useful for various types of effective Hamiltonians to describe Dirac/Weyl semimetals.

{\it Physical example 3: lower-dimensional materials.}---In the case of two-dimensional Dirac fermions, as described by $d=2$ in the formula (\ref{eq:d_dep}), promising systems are Dirac electrons living on graphene and surface states on topological insulators~\cite{Ishikawa:2020icy}.
In particular, carbon nanotubes can be regarded as a platform of the electronic Casimir effect with the PBC~\cite{Bellucci:2009jr,Bellucci:2009hh}.

\section{Conclusion}

In this Letter, we focused on the Lifshitz formulas for the Dirac fermions at finite chemical potential and its application.
Similarly, the formulas for bosonic fields, such as charged scalar and gauge fields, will be straightforward.
If bosonic chemical potentials are experimentally controllable, the corresponding formulas will be significant tools.

\section*{Acknowledgments}

This work was supported by the Japan Society for the Promotion of Science (JSPS) KAKENHI (Grants No. JP20K14476, JP24K07034, JP24K17054, JP24K17059).

  \setcounter{section}{0}
  \setcounter{equation}{0}
  \setcounter{figure}{0}
  \renewcommand{\theequation}{A\arabic{equation}}
  \renewcommand{\thesection}{A\arabic{section}}
  \renewcommand{\thefigure}{A\arabic{figure}}

\appendix

\section{Lattice regularization approach} \label{App:lattice}
Here, we explain the lattice regularization approach~\cite{Actor:1999nb,Pawellek:2013sda,Ishikawa:2020ezm,Ishikawa:2020icy,Nakayama:2022ild,Nakata:2022pen,Mandlecha:2022cll,Nakayama:2022fvh,Swingle:2022vie,Nakata:2023keh,Flores:2023whr,Nakayama:2023zvm,Beenakker:2024yhq,Fujii:2024fzy} for calculating the Casimir energy.
In this approach, the Casimir energy is defined as the difference between the zero-point energies, $E^{\rm sum}_{0}$ in finite $L_z$ and $E^{\rm int}_{0}$ in infinite volume, where these two quantities are regularized by the lattice cutoff characterized by a lattice spacing $a=L_i/N_i$ ($i=x,y,z$) with the number of lattices $N_i$.
At a nonzero chemical potential $\mu$, the Casimir energy per unit area is~\cite{Fujii:2024fzy}
\begin{align}
E^{\rm Lat}_{\rm Cas} \equiv&E^{\rm sum}_{0}-E^{\rm int}_{0} \label{eq:ECaslat}, \\
E^{\rm sum}_0=&-\frac{2}{a^3} \int_{\rm BZ}\frac{d(ak_x)d(ak_y)}{(2\pi)^2}\notag \\
&\times \sum^{\rm BZ}_{n} \left[\frac{1}{2}a|E^{\rm Lat}_{n}-\mu|+\frac{1}{2}a|E^{\rm Lat}_{n}+\mu|\right],
\\
E^{\rm int}_{0}=&-\frac{2}{a^3} \int_{\rm BZ}\frac{d(ak_x)d(ak_y)d(ak_z)}{(2\pi)^3} \notag\\
&\times N_z\left[\frac{1}{2}a|E^{\rm Lat}-\mu|+\frac{1}{2}a|E^{\rm Lat}+\mu|\right].
\end{align}
The overall factor $-2$ comes from the minus sign of the fermion zero-point energy and the two spin degrees of freedom.
The form of $\frac{1}{2}|E-\mu|+\frac{1}{2}|E+\mu|$ is well known in the context of relativistic quantum field theory at finite density, where $\frac{1}{2}$ is the zero-point energy factor.
$E^{\rm Lat}_n$ and $E^{\rm Lat}$ are the energy eigenvalues of lattice fermions at $\mu=0$.
The choice of lattice fermions is optional, and
for example, we can use 
\begin{align}
&E^{\rm Lat}=\sqrt{M^2+ \frac{1}{a^2}\sum_{i=x,y,z}(2-2\cos ak_i)}, \\
&E^{\rm Lat}_n =\sqrt{M^2+ \frac{1}{a^2}\left[\sum_{i=x,y}(2-2\cos ak_i)+\left(2-2\cos \frac{2n\pi}{N_z}\right)\right]},
\end{align}
as a massive-Dirac-like lattice fermion.
The momentum integration and summation ranges are taken within the first Brillouin zone (BZ): $0\leq ak_i<2\pi$  (or equivalently $-\pi \leq ak_i< \pi$) and $n=0,1,\dots,N_z-1$ (or equivalently $n=1,2,\dots,N_z$) for the PBCs.

Finally, by taking the continuum limit $a\to0$, we can obtain the Casimir energy in the continuum spacetime:
\begin{align}
E_{\rm Cas} = \lim_{a\to 0} E^{\rm Lat}_{\rm Cas}.
\end{align}
This equation holds if we adopt an appropriate lattice regularization that correctly captures the property in the continuum theory.

\bibliographystyle{apsrev4-1}
\bibliography{ref}

  \setcounter{section}{0}
  \setcounter{equation}{0}
  \setcounter{figure}{0}
  \renewcommand{\theequation}{S\arabic{equation}}
  \renewcommand{\thesection}{S\arabic{section}}
  \renewcommand{\thefigure}{S\arabic{figure}}



\end{document}